# A universal route for the formation of massive star clusters in giant molecular clouds

Corey S. Howard[1,*], Ralph E. Pudritz[1,2], & William E. Harris[1]

[1]Department of Physics & Astronomy, McMaster University, 1280 Main Street West, Hamilton, ON, L8S 4L8, Canada, howardcs@mcmaster.ca
[2]Origins Institute, McMaster University, 1280 Main Street West, Hamilton, ON, L8S 4M1, Canada

**Young massive star clusters (YMCs, with $M \geq 10^4 \, M_\odot$) are proposed modern-day analogues of the globular clusters (GCs) that were products of extreme star formation in the early universe[1-4]. The exact conditions and mechanisms under which YMCs form remain unknown[4,5] – a fact further complicated by the extreme radiation fields produced by their numerous massive young stars[6-9]. Here we show that GC-sized clusters are naturally produced in radiation-hydrodynamic simulations of isolated $10^7 \, M_\odot$ Giant Molecular Clouds (GMCs) with properties typical of the local universe, even under the influence of radiative feedback. In all cases, these massive clusters grow to GC-level masses within 5 Myr via a roughly equal combination of filamentary gas accretion and mergers with several less massive clusters. Lowering the heavy-element abundance of the GMC by a factor of 10 reduces the opacity of the gas to radiation and better represents the high-redshift formation conditions of GCs[10,11]. This results in higher gas accretion leading to a mass increase of the largest cluster by a factor of ~4. When combined with simulations of less massive GMCs[12] ($10^{4-6} \, M_\odot$), a clear relation emerges between the maximum YMC mass and the mass of the host GMC. Our results demonstrate that YMCs, and potentially GCs, are a simple extension of local cluster formation to more massive clouds and do not require suggested exotic formation scenarios[14-16].**

    Star clusters are the cradles of star formation and grow within Giant Molecular Clouds (GMCs) – large collections of turbulent molecular gas and dust with masses of $10^{4-7} \, M_\odot$ and scale sizes typically 10 – 200 parsecs[17,18]. Most star clusters forming in lower mass clouds ($10^{4-6} \, M_\odot$) in the local Universe are relatively low mass ($<10^4 \, M_\odot$)[19,20] but, while rare, there are much higher-mass examples nearby (called Young Massive Clusters or YMCs, $>10^4 \, M_\odot$). The infrequency of local YMCs, in combination with their short < 10 Myr timescale of formation, makes their formation conditions and mechanisms uncertain. In particular, it is debated whether YMCs form from the hierarchical collapse of a single high density region of molecular gas, or whether they are assembled via the conglomeration of several distinct subclusters[2,5].

    Interpretations are further complicated by the extreme radiation fields produced by the many hot O stars in YMCs, which heat and ionize the gas surrounding the cluster and act to suppress star formation[6,9,12]. The associated radiation pressure on dust grains can help remove the natal gas[7,8]. These effects, deemed "radiative feedback", decrease the star formation efficiency within a GMC. Most previous studies of radiative feedback, however, focus on low-mass cluster formation[6,9], so the degree to which feedback shapes the formation of YMCs is not well known.

    At the high-mass end of star clusters but at the opposite end of the age scale are the globular clusters (GCs) – relics from an epoch of extreme star formation in the early universe (redshifts z > 2). GCs have present-day masses $\sim 10^{4-7} \, M_\odot$ and cover a wide range of metallicities (heavy-element

abundances) that are typically subsolar (-2.5 < log Z/Z$_\odot$ < 0)[21,22]. Direct observation of GC formation at high redshift is only now becoming possible[23-25]. This, in combination with evidence for multiple stellar populations in GCs, has resulted in numerous scenarios that invoke special conditions in the early universe to explain their formation[14-16]. YMCs are present-day analogs of GCs[3], so understanding YMC formation may provide deep insights into the origin of GCs.

We study the formation and assembly of YMCs during the first crucial ~5 Myr of their evolution through radiation-hydrodynamic (RHD) simulations of isolated, turbulent $10^7$ M$_\odot$ GMCs with physical properties typical of those found in our Galaxy. Clouds of this mass are indeed present in the local universe, but are thought to be more abundant during the much more gas-rich early universe when GCs formed[1]. The imposed turbulence produces a system of dynamically evolving filaments, in which clusters form (Supplementary Videos 1-4). We perform two simulations: one at Solar metallicity (Z$_\odot$), and one at a tenth Solar metallicity (0.1 Z$_\odot$) that is closer to matching the early universe. Star clusters are represented via localized sink particles and their stellar content is prescribed by a subgrid model. One main parameter of the subgrid model is the density threshold for cluster formation which we take to be $10^4$ cm$^{-3}$ based on the observational divide between starless and star-forming clumps in the local Milky Way[19]. We have tested the effects of varying this parameter up to $10^6$ cm$^{-3}$ and find that the final total mass of the most massive cluster is little affected (see Methods and Extended Data Fig. 1). Clusters grow both by accreting gas from their host filaments and by merging with other clusters. A raytracing radiative-transfer scheme is used for heating, solving the ionization state of the gas, and inducing radiation pressure.

This work represents the first time the detailed evolutionary history of massive clusters in a $10^7$ M$_\odot$ GMC with the inclusion of ionization feedback and radiation pressure has been studied. We do not include the effects of stellar winds which have been shown to reduce the star formation in young star-forming regions[26]. We also do not supernovae (SNe) feedback in our simulation but have verified that SNe would not significantly alter our results over the timescales considered (see Methods).

Here we focus on the formation and evolution of the most massive cluster in each simulation, referred to as the YMC hereafter. The simulations show (Supplementary Videos 1-4) that clusters are born within filaments and move with the filamentary gas flows. The YMCs form in high column density filaments at 1.54 and 0.84 Myr for the Z$_\odot$ and 0.1 Z$_\odot$ GMCs (Fig. 1a and 1b). Both YMCs continue to gain mass by gas inflow from their host filaments, but they also grow by several merging events. The smaller clusters that they capture originate as distant as 21 pc (Z$_\odot$) and 41 pc (0.1 Z$_\odot$) away from the formation location of the YMC, though the average separations are 15 pc and 19 pc. The entire GMC environment, therefore, needs to be considered when tracing YMC formation. There are a total of 229 and 146 cluster particles at the end of the simulations, resembling the subclustered regions and star-forming clumps in 30 Doradus[27].

Examining the YMC histories reveals key details about their growth (Fig. 2). The Z$_\odot$ cluster undergoes 9 mergers that are roughly equally spaced throughout its lifetime. Conversely, the 0.1 Z$_\odot$ cluster participates in 23 mergers with most occurring after 3 Myr. Most mergers are with smaller clusters; the average mass ratio between the captured cluster and the YMC at time of merger is 8.2% and 9.9%. Mergers occur primarily between clusters moving within the filaments. The small clusters also grow by gas accretion, but rarely undergo mergers themselves before being absorbed by the YMC.

The final total masses of the YMCs at the end of the 5 Myr run are 2.84x$10^5$ M$_\odot$ (for Z$_\odot$) and 1.54x$10^6$ M$_\odot$ (for 0.1 Z$_\odot$) including both their stellar and gaseous components (see Methods). Of this mass, 50% and 46% was obtained through mergers. Their final stellar masses excluding gas are 2.12x$10^5$ M$_\odot$ and 8.69x$10^5$ M$_\odot$, placing them well within the range of observed YMCs and GCs. The stellar mass quoted here represents the total mass of the stars formed in the YMC. The bound component is typically less than the total stellar mass by a fraction that depends on the cluster's local environment[1].

The numbers of mergers are stochastically driven, but the factor of ~4 difference in total mass between the two YMCs is a physical effect of their metallicity difference. At subsolar metallicity, the radiation produced by the young stars exerts less pressure on the surrounding gas because the opacity (mainly contributed by dust grains) is lower, allowing the YMC to continue accreting gas throughout its entire history.

In Figure 3, we show the net gas accretion rate into a 5 pc radius sphere centered on the YMC. For the first ~2 Myr, both clusters show a moderate flow inward, but afterwards, the flow rate oscillates as the particle moves through turbulent gas of varying density (Supplemental Videos 1-4). At ~3.5 Myr, the radiative pressure produced by the YMC in the Solar-metallicity GMC grows sufficiently large to clear its surroundings, creating a large radiatively driven outflow bubble followed by a period of quiescence. The cluster grows only by mergers past after this time (Extended Data Fig. 2 & 3). By contrast, the YMC in the low-metallicity GMC shows no such radiatively-driven bubble and gas accretion continues to the end of the simulation (Extended Data Fig. 4).

To isolate which particular feedback mechanism is responsible for the different final YMC masses, we ran simulations that include ionization and heating but neglect radiation pressure. For the $Z_\odot$ GMC, excluding radiation pressure results in a net star formation efficiency (SFE) that is nearly indistinguishable from a purely hydrodynamical simulation that includes no form of feedback (Extended Data Fig. 5). Therefore, as also suggested by other authors[7,8], radiation pressure is the most important form of stellar feedback during the early stages of YMC formation whereby they can suppress their own gas accretion. The SFEs presented in Extended Data Fig. 4 are higher than the observed values, possibly because feedback from stellar winds or magnetic fields are not included.

Lastly, we combine the above results for $10^7$ $M_\odot$ GMCs with our earlier simulations for lower mass GMCs ($10^{4-6}$ $M_\odot$) to uncover the continuity between YMCs and their lower-mass analogues. Plotting the stellar mass of the most massive cluster ($M_{max}$) that forms at a given GMC mass ($M_{GMC}$) reveals a clear power-law relation (Fig. 4). For $Z_\odot$ molecular clouds in RHD simulations, we find $M_{max} \propto M_{GMC}^{0.78}$, whereas purely hydrodynamical simulations give $M_{max} \propto M_{GMC}^{0.92}$. This scaling, over 3 orders of magnitude in GMC mass, indicates that YMCs are a natural extension of low-mass cluster formation. Since the initial cloud properties are typical of GMCs in the local Universe, no special conditions, other than sufficiently massive GMCs, are required for their formation. We note that the normalization of this relation may include an environmental, and therefore redshift, dependence. This is supported by recent work suggesting that the most massive cluster depends on the SFE and the bound stellar fraction, both of which depend on the local environment[28].

Our work draws out the importance of both continued filamentary gas accretion as well as multiple mergers and hierarchical growth in the formation history of massive star clusters, from the very earliest times in their evolution. The progenitor clusters that combine to produce the final YMC themselves started at widely scattered locations in the entire GMC, which opens up the possibility for combining different chemical evolution histories[29]. The power-law relation between YMC and cloud mass is shallower for radiative feedback effects in the current universe. Evidently, low metallicity environments typical of the early universe naturally build more massive clusters due to reduced opacity and feedback which also has implications for galaxy formation[30]. The extension of these concepts further towards the GC regime promises to lead to a better understanding of their formation.

Formation Simulations. *Astrophys. J.* **804**, 18 (2015)

**Supplementary information** available online

**Acknowledgments** This research was supported by financial support provided by the Natural Sciences and Engineering Research Council (NSERC) through a Postgraduate scholarship and Discovery Grant numbers #. Computations were performed on the gpc supercomputer at the SciNet HPC Consortium. SciNet is funded by: the Canada Foundation for Innovation under the auspices of Compute Canada; the Government of Ontario; Ontario Research Fund - Research Excellence; and the University of Toronto.

**Author Contributions** CSH carried out the simulations, completed the data analysis and figure production, and led the manuscript preparation. WEH and REP contributed to the interpretation and presentation of the data, and helped with the production of the manuscript.

**Author Information** Reprints and permissions information available at www.nature.com/reprints. The authors declare no competing financial interests. Readers are welcome to comment on the online version of the paper. All correspondence and requests for materials should be addressed to CSH (howardcs@mcmaster.ca).

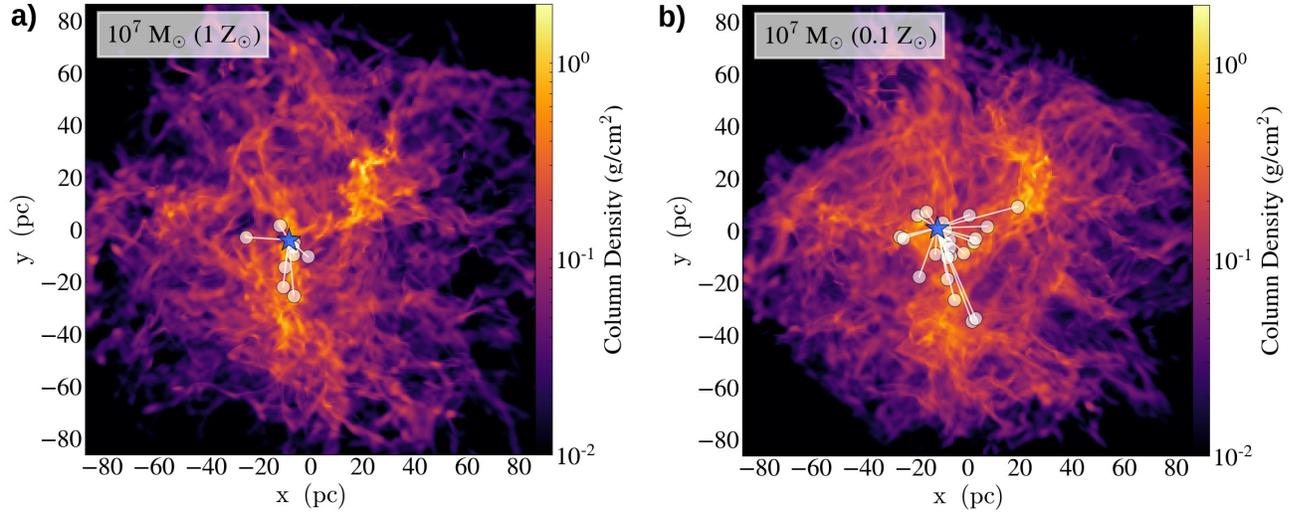

Figure 1: **The formation of the most massive star cluster and its merging partners. (a):** A column density projection of the $Z_\odot$ GMC along the z-axis at t = 1.54 Myr, corresponding to the moment when what will ultimately become the most massive cluster forms. The location of the YMC is shown as the blue star. This cluster initially appears in a high column density filament. The white circles represent the formation locations of the clusters that will eventually merge with the YMC (though these clusters are not necessarily at this location at this time). **(b):** The same Figure for the 0.1 $Z_\odot$ GMC.

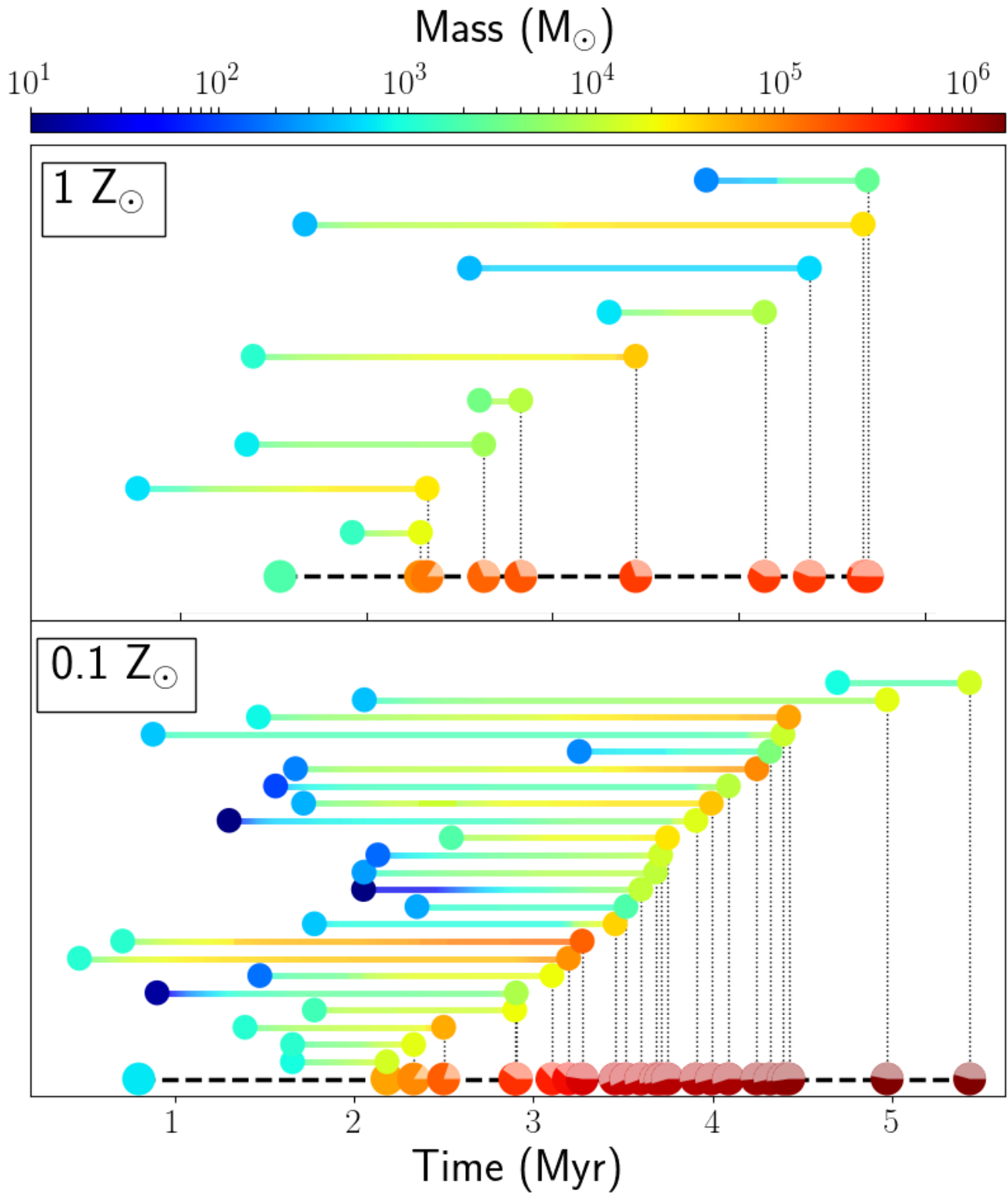

Figure 2: **The accretion-merger histories of the YMCs.** The mass evolution of the $Z_\odot$ YMC (top) or $0.1\,Z_\odot$ YMC (bottom) is shown as the large circle at the bottom of each panel. The symbol color represents its mass at the time of each merger. The lighter shaded region on each YMC marker represents the fraction of mass obtained via mergers up to that point. Smaller circles above the YMC show the masses of its merging partners at the time of formation and at the time of merging, while the colored connecting lines trace their mass growth due to gas accretion.

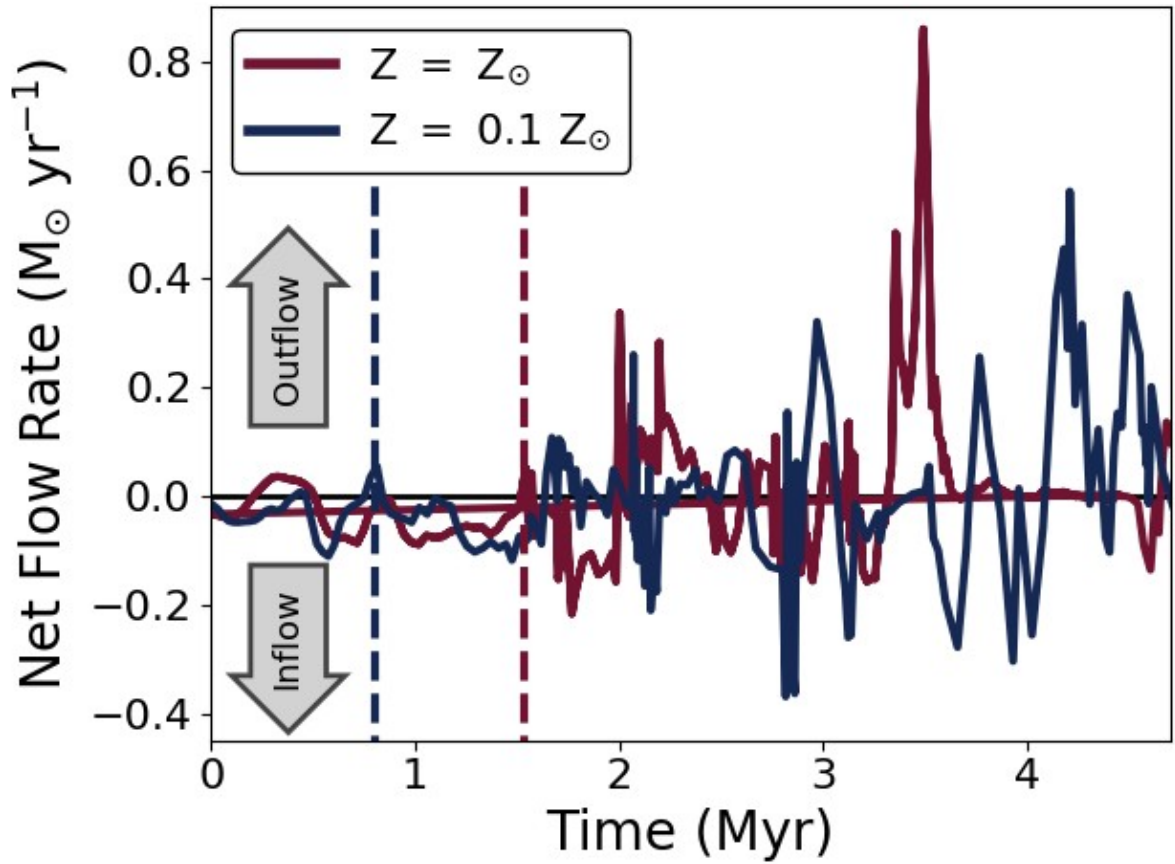

Figure 3: **The gaseous flow rate in the vicinity of the YMCs.** The net mass flow rate into a sphere of 5 pc radius that is centered on the YMC is shown, where the sphere follows the cluster as it evolves. The vertical dashed lines show the time each cluster formed. After the strong radiation pressure-driven outflow at t ~ 3.3 Myr, the $Z_\odot$ YMC does not accrete further.

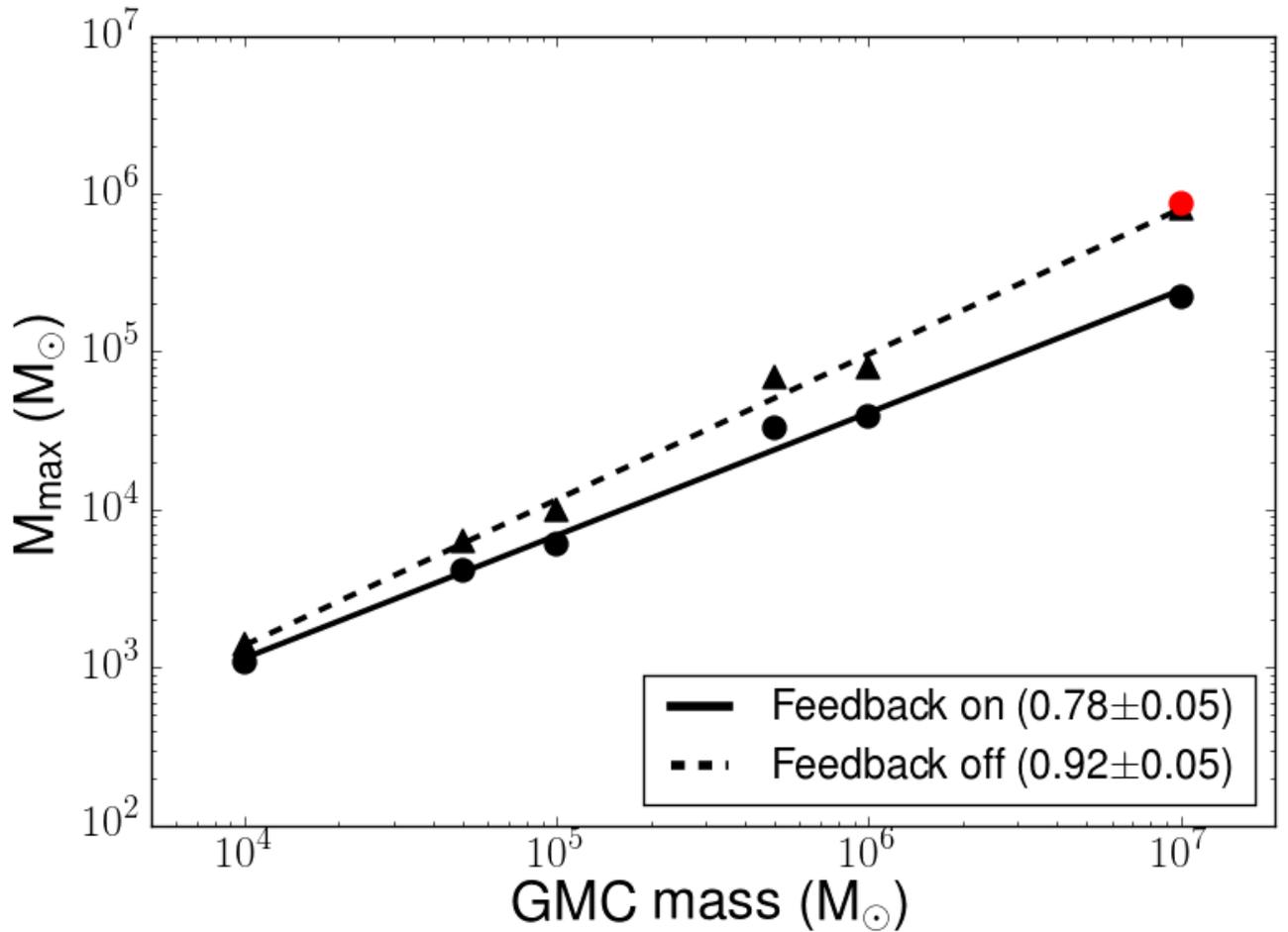

Figure 4: **The maximum cluster mass produced by a given GMC mass**. The stellar mass of the most massive cluster formed in a GMC is plotted versus GMC mass. Linear regressions with their resulting slopes and uncertainties are shown in the legend. Black points refer to $Z_\odot$ simulations and the red point shows the lower metallicity 0.1 $Z_\odot$ simulation described in the text. Feedback results in a significantly shallower slope compared to the purely hydrodynamic simulations for the solar metallicity case (but not subsolar metallicity which follows the purely HD result).

Methods

**The FLASH code.** We use version 2.5 of the Adaptive Mesh Refinement (AMR) code FLASH[31] which solves the compressible gas dynamics equations on a Eulerian grid. The grid structure and refinement is handled via the block-structured PARAMESH[32] package. A directionally split Piecewise-Parabolic Method (PPM) is used for the hydrodynamic calculations and self-gravity is included through the use of a multigrid Poisson solver.

For the gas, we use a perfect-gas equation of state with $\gamma = 5/3$ and a mean molecular weight of 2.14. Gas cooling via $H_2$ dissociation, molecular line emission, and gas-dust interactions are included[33] with the cooling rates for the latter two processes coming from [34] and [35].

**Sink particle implementation.** Sink particles – Lagrangian particles used to replace dense, gravitationally bound, and collapsing regions of gas – are used to represent star clusters. The specific implementation in FLASH is described in [36]. In order for a sink particle to form, several stringent conditions must be met. A region of gas must: 1) Be above an adopted density threshold. 2) Be at the highest level of refinement. 3) Have a negative local velocity divergence (ie. converging flow). 4) Be located at a local gravitational minimum. 5) Be Jeans unstable. 6) Be gravitationally bound. 7) Not be within the accretion radius (defined as 2.5 cells at the highest level of refinement) of another particle. When these conditions are met, a particle is formed and is given the mass of all the gas within its radius. The particle is then free to move through the simulation volume and interact gravitationally with its surroundings.

We allow for sink particle mergers under certain circumstances. In order to merge, particles must be separated by less than a particle radius, they must have a negative relative velocity, and be gravitationally bound to one another. When particles merge, their masses are combined, the new particle is placed at the system's center of mass, and it is assigned the center of mass velocity.

**Subgrid model for star formation.** We have developed a custom subgrid model to prescribe how star formation proceeds within the cluster sink particles[37]. We adopt a density threshold for cluster formation of $10^4$ cm$^{-3}$. This is the observational divide between starless and star-forming clumps[19], and several theoretical models for cluster formation predict a similar threshold[38,39]. When the cluster initially forms, we assume its mass is solely in gas – deemed the "reservoir" – that has not yet been used to form stars. We convert the reservoir gas to stars by sampling a Chabrier[40] Initial Mass Function (IMF) with an efficiency of 20% of the reservoir mass per freefall time. The freefall time is 0.36 Myr, corresponding to gas at $10^4$ cm$^{-3}$. This specific value for the star formation efficiency was chosen to be consistent with observations of local star-forming clumps[19] in the Milky Way. The IMF is sampled every tenth of a freefall time to allow the cluster's stellar mass to grow smoothly with time. The masses of all stars formed in each cluster are recorded, and any gas that is accreted by the cluster is added to the reservoir and used for future star formation.

The total luminosity and total ionizing luminosity at any time t for each cluster are calculated based on their stellar populations. Metallicity-dependent analytic fits from [41] determine the main-sequence luminosity and temperature (L, $T_{eff}$) of each star from their mass. We assume the stars have the same metallicity as their host GMC (ie. either 1 $Z_\odot$ or 0.1 $Z_\odot$). We do not include protostellar evolution. The ionizing UV luminosity is calculated by assuming each star radiates as a blackbody and integrating the Planck function for energies greater than 13.6 eV. Summing over all stars in the cluster gives the total luminosity and ionizing luminosity which are used by the radiative transfer scheme.

To reduce the computational time of the radiative transfer scheme, which accounts for a large portion of total computational expense, we introduce a mass threshold of $10^4$ $M_\odot$ below which clusters do not radiate. These small clusters continue to form stars, accrete gas, and participate in gravitational interactions, but they are too small to produce many O stars and thus are not considered in the radiative transfer calculations. We have verified that <10% of the total luminosity in the simulation is contained in clusters below the threshold.

**Radiative Transfer.** We use a hybrid-characteristics raytracing scheme, developed by [42] and expanded for astrophysical use by [43], to treat radiative transfer. This scheme employs the block based structure of the PARAMESH grid to combine a short-characteristic method (used across blocks) and a long-characteristic method (used within blocks) to make an efficient and parallel radiative transfer solver. The scheme follows the propagation of both ionizing and non-ionizing radiation.

The flux of ionizing photons, taken to be photons with energies > 13.6 eV, is used by the DORIC routines[44] to calculate the ionization state of the gas by considering photoionization and "case B" radiative recombinations. For simplicity, the DORIC routines assume that hydrogen is the only gas species. Non-ionizing radiation is used as a heating source when calculating the temperature of the gas. We adopt the temperature dependent Planck mean opacities from [45] for the non-ionizing radiation. These opacities were calculated for mixtures of gas, silicates, ices, and organics with abundances typical of the ISM.

We have expanded the raytracer to include radiation pressure. The radiative force per unit mass ($F$) exerted on a cell separated from a source of luminosity ($L$) by a distance $r$ is,

$$F = \frac{\kappa L}{c} \frac{e^{-\tau_{uv}}}{4\pi r^2}$$

where $\kappa$ is the opacity to ionizing radiation, and $\tau_{UV}$ is the optical depth between the source and the cell. We use a single UV opacity[46] for radiation pressure with a value of 775 cm$^2$ g$^{-1}$. This opacity is scaled by the neutral fraction of the gas, so that fully ionized regions have zero opacity. We do not include photon scattering or the emission of reprocessed infrared radiation.

We model gas of different metallicity by assuming that the opacity – both to ionizing and non-ionizing radiation – scales linearly with the heavy-element abundance Z. Since the main source of opacity is contributed by dust grains, this inherently assumes that the gas-to-dust ratio also scales linearly with metallicity. This has been shown to be valid down to 0.1 $Z_\odot$ through studies of the ISM in other galaxies[47].

**GMC initial conditions.** Our model GMCs are initially spherical with radii of 77 pc and are embedded in a cubic box with a side length of 173 pc. The highest resolution is 0.6 pc and outflow boundary conditions are used for the domain edges. Matter may flow out of the box, so the total mass in the simulation volume is not conserved.

The initial density profile ρ(r) is uniform in the central half of the cloud and decreases as ρ ~ $r^{-3/2}$ in the outer half. A quadratic fit is applied at the transition region to ensure a continuous and smooth density distribution. The density outside the GMC is 100 times less than the density at the cloud surface and the temperature is chosen such that the GMC and external medium are in pressure equilibrium. The temperature inside the cloud is initially 10 K.

We overlay each GMC with an initial Burgers turbulent velocity spectrum, as in [48]. The turbulent spectrum contains a natural (random) mixture of solenoidal and compressive modes. The turbulence is not driven as the simulation evolves. The strength of the turbulence is determined by choosing the initial virial parameter $\alpha = 2\,E_K/E_{grav}$. We set $\alpha(t=0) = 3$ (i.e. initially unbound) because, as shown in [49], it results in low SFEs and the cloud quickly becomes virialized since the turbulence is not continuously driven. The same velocity spectrum is applied to each cloud in order to isolate the effects of radiative feedback and metallicity.

All models are evolved for ~5 Myr. The simulations are ended at this time for two reasons. Firstly, the computational expense increases as the simulations evolve because the dynamic range of the density structure increases dramatically, and the number of radiating clusters grows with time. Secondly, we do not include the effects of supernovae so we stop the simulations before the supernovae phase is expected to significantly alter the course of the simulation.

We have verified that neglecting supernovae feedback for the 5 Myr of evolution presented here is a reasonable assumption. The majority of O stars have lifetimes that exceed the length of our simulation, but very massive stars (> 60 $M_\odot$) can enter the supernovae phase within ~3.5 Myr[50]. Based on the stellar populations in the YMCs, the first supernovae from stars of this mass are expected at 5.2 and 4.5 Myr for the $Z_\odot$ and 0.1 $Z_\odot$ simulations. For the first case, this occurs after the simulation has ended. For the second case, the YMC has already obtained ~90% of its mass by this time. Therefore, even if the first supernovae completely halt star formation and further gas accretion, the results of this paper will not be significantly affected.

**Resolution Tests** We have tested the robustness of our results by varying the resolution (and threshold density for cluster formation, see below) in our 0.1 $Z_\odot$ simulations. To reduce the overall computational time required for these tests, we have chosen to complete HD simulations since the radiative transfer routine is the most expensive physics module in our code. In Extended Data Fig. 1, we show the total mass of the YMCs in solid lines and the stellar mass in dashed lines. Since our results focus heavily on the YMCs mass and the role of merging in their growth, comparing these properties in different test cases is sufficient for showing that our results are not a strong function of our adopted parameters.

Lowering the resolution by a factor of two results in a modest increase in the total mass from $1.52 \times 10^6\,M_\odot$ in the fiducial case to $1.76 \times 10^6\,M_\odot$. The stellar mass is also increased by approximately the same amount. This comparison, as well as those discussed below, is made at 4.68 Myr corresponding to the end of the increased density threshold simulation. Both YMCs also form at similar times (0.71 and 0.54 Myr).

**Formation Density Threshold Tests** Increasing the threshold density for cluster formation from $10^4$ to $10^5$ cm$^{-3}$ does not change the final total mass of the YMC ($1.52 \times 10^6\,M_\odot$ vs $1.51 \times 10^6\,M_\odot$). Because the gas needs to collapse to higher densities to form clusters, the YMC forms later at 1.45 Myr for the increased threshold. But, the YMC quickly grows in mass to meet the fiducial case because the cluster particles accrete gas from a fixed radius, and since the gas within that radius has reached higher densities, the mass increases faster. The stellar mass in the case of an increased density threshold is $1.19 \times 10^6\,M_\odot$ compared to $7.99 \times 10^5\,M_\odot$ for the fiducial simulation. This difference of a factor of ~1.5 is due to the subgrid model for star formation described above. To convert the reservoir gas to stars, we adopt a star formation efficiency per freefall time where the freefall time is calculated for the adopted formation density threshold. Since an increased density results in a smaller freefall time, the overall rate of star formation is increased. The average SFR over the entire course of the YMCs evolution is increased by a factor of ~3.5 relative to the fiducial case. This is consistent with observations that show a positive relation between the SFR in a GMC and the fraction of dense gas (> $10^4$ cm$^{-3}$) contained within it[51] since a simulation with a higher formation threshold will have a larger fraction of dense gas.

Increasing the density threshold even further to $10^6$ cm$^{-3}$ reinforces these trends. The total YMC

mass decreases from $1.52 \times 10^6 \, M_\odot$ to $1.35 \times 10^6 \, M_\odot$. The stellar mass is increased by a factor of 1.5 compared to the fiducial case despite the YMC forming at 1.92 Myr. This results in an **average SFR** that is higher by a factor of ~15.

While there are minor differences in the mass of the YMCs when the density threshold is changed, we have shown through these numerical tests that a YMC can indeed form within 5 Myr. Since the clusters have fixed physical sizes from which they accrete their gas, increasing the density threshold for formation will delay their appearance but they will grow faster by accreting higher density gas. This results in YMCs with similar final masses, regardless of the adopted threshold. We believe this is consistent with observations that show the SFR in a given GMC that is actively star-forming scales directly with the amount of dense gas contained within it.

**YMC Merging in our Numerical Tests** Since our results focus heavily on the role of mergers in the growth of YMCs, we have also examined the merging histories in our various numerical test simulations. The final fraction of the YMCs mass obtained via mergers (and the total number of mergers) for the fiducial, lower resolution, 10x formation threshold, and 100x formation threshold simulations are 59% (29), 34% (8), and 31% (8), 18% (2). In the latter three cases, the total number of merging events involving the YMC is most likely lower because fewer clusters are formed overall. However, the total fractional mass obtained from mergers is still significant. We will examine how the global properties of our GMCs and clusters (ie. gas dynamics, formation efficiencies, cluster mass distributions, etc.) are affected by changing our simulation parameters in a future work.

**Flow rate calculation.** To calculate the net flow of gas into a 5 pc spherical region centered on the YMC, we first extract a 3D spherical region from the FLASH grid. We chose a 5 pc radius because it encompasses the particle accretion radius of 1.5 pc and is sufficiently small for the surrounding gas to be bound to the YMC. The YT analysis toolkit[52] is used for the extraction. To calculate the flow of gas into the sphere, we compute the total mass in the sphere ($M_i$) at each time ($t_i$). The flow rate ($\dot{M}_i$) at any given time is given by,

$$\dot{M}_i = (M_i - M_{i-1}) / (t_i - t_{i-1})$$

and is plotted in Fig. 3.

We produce further visualizations that show the spatial variations of the density and flow rate across the surface of the 5 pc sphere (Extended Data Figure 2, Supplementary Videos 5&6). To make these images, we first extract a 2D spherical surface using the marching cube algorithm implemented in YT. The gas density and velocity are known at every point on the surface. To calculate the flow across the surface (in units of mass/time/area), we take the dot product of the velocity and the radial unit vector and multiply the resulting scalar by the density. To visualize the density and mass flow rate across the sphere in a 2d image, we apply a Hammer-projection equal-area map. This projection was chosen for its area conserving properties and reduced distortion in the polar regions. We align the projection such that the poles are along the z-direction.

**Extended Data:**

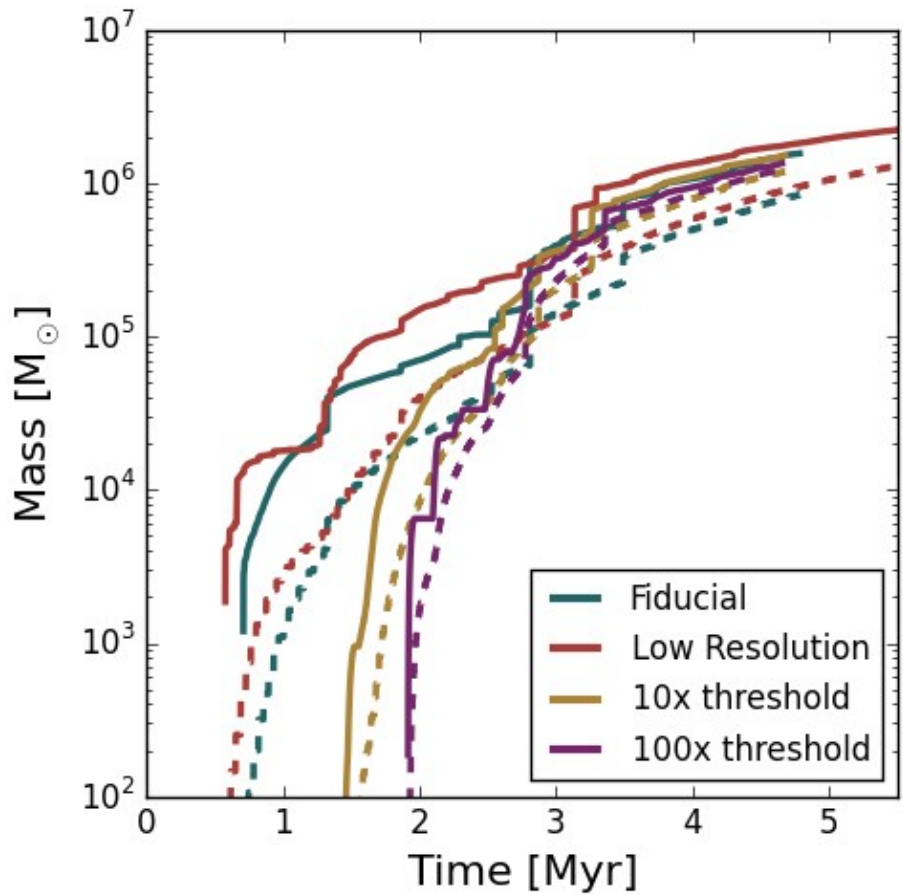

Extended Data Figure 1: **Tests varying the resolution and cluster formation density threshold.** The total mass (solid lines) and the stellar mass (dashed lines) of the YMC are shown as a function of time for purely HD simulations that vary the resolution and density threshold for cluster formation. The final masses of the YMCs are not strongly altered by numerical changes.

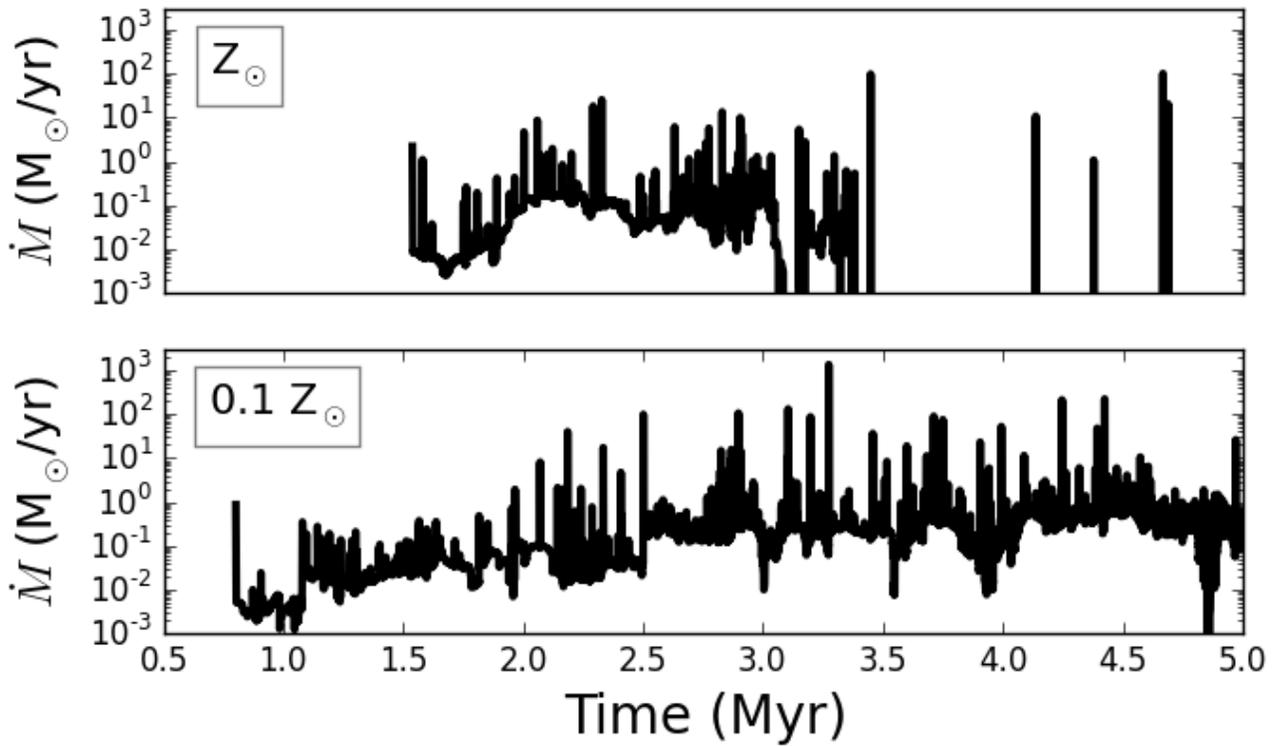

Extended Data Figure 2: **Total cluster growth rates.** The accretion rates, including both gas and cluster mergers, of the $Z_\odot$ YMC (top) and the 0.1 $Z_\odot$ YMC (bottom). After the radiatively driven outflow at ~3.3 Myr in the $Z_\odot$ case, the YMC grows only via mergers, as shown by the discrete accretion events at late times (see Fig. 2). The 0.1 $Z_\odot$ YMC accretes both gas and clusters throughout its entire evolution.

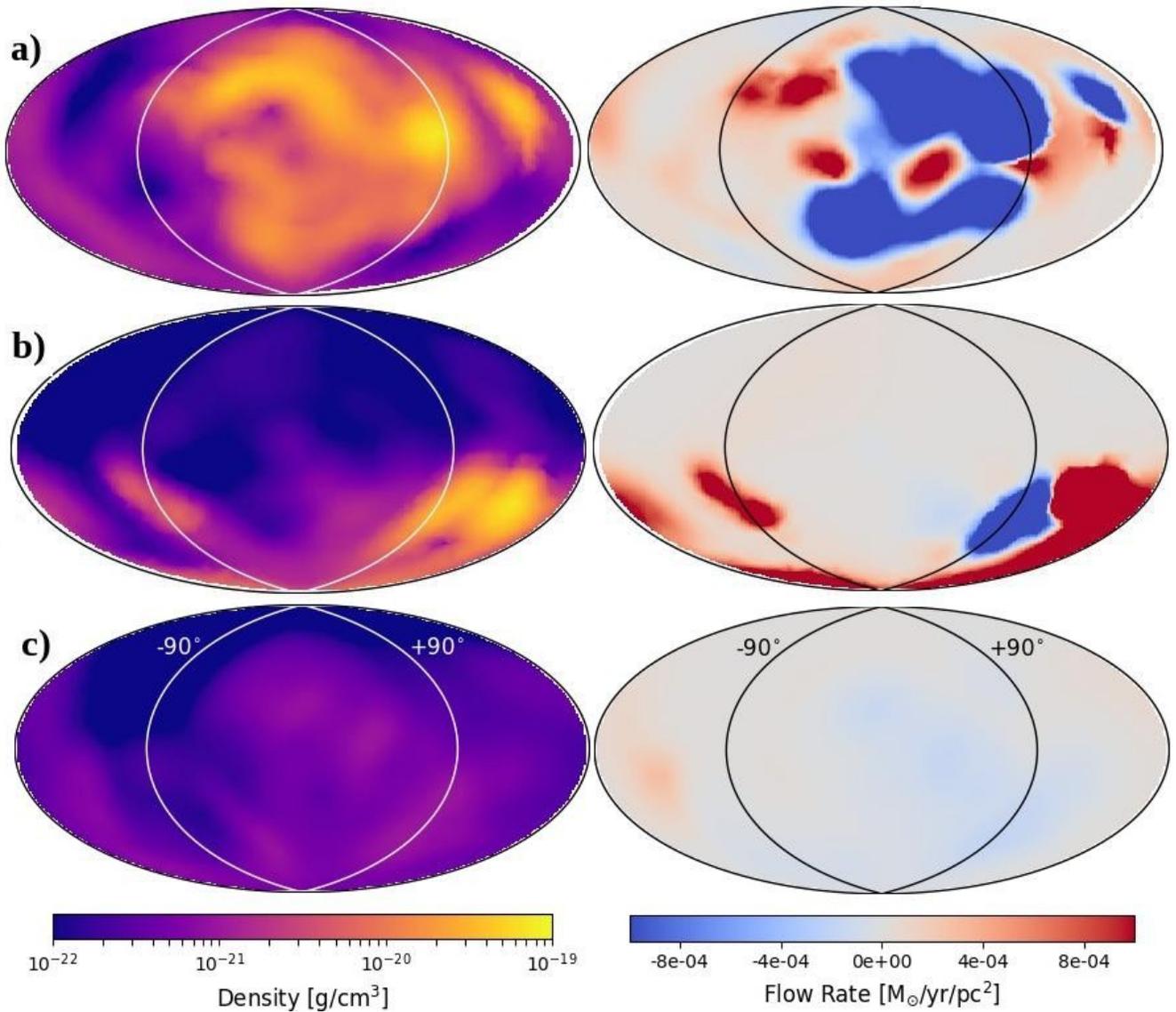

Extended Data Figure 3: **The local conditions surrounding the $Z_\odot$ YMC. (a)** The density (left) and mass flow rate (right) at the surface of a 5 pc sphere surrounding the YMC. In the mass flow plots, blue and red indicate inflow and outflow. Data is plotted at 3.20 Myr, just before the YMC clears its surroundings via the radiatively-driven outflow. The solid lines indicate angles of ±90º (ie. the limits of the front-facing side of the sphere).**(b)** The data plotted during the radiatively-driven outflow at 3.53 Myr. **(c)** The YMC's local conditions after the outflow, plotted at 3.69 Myr. There is little change after this point.

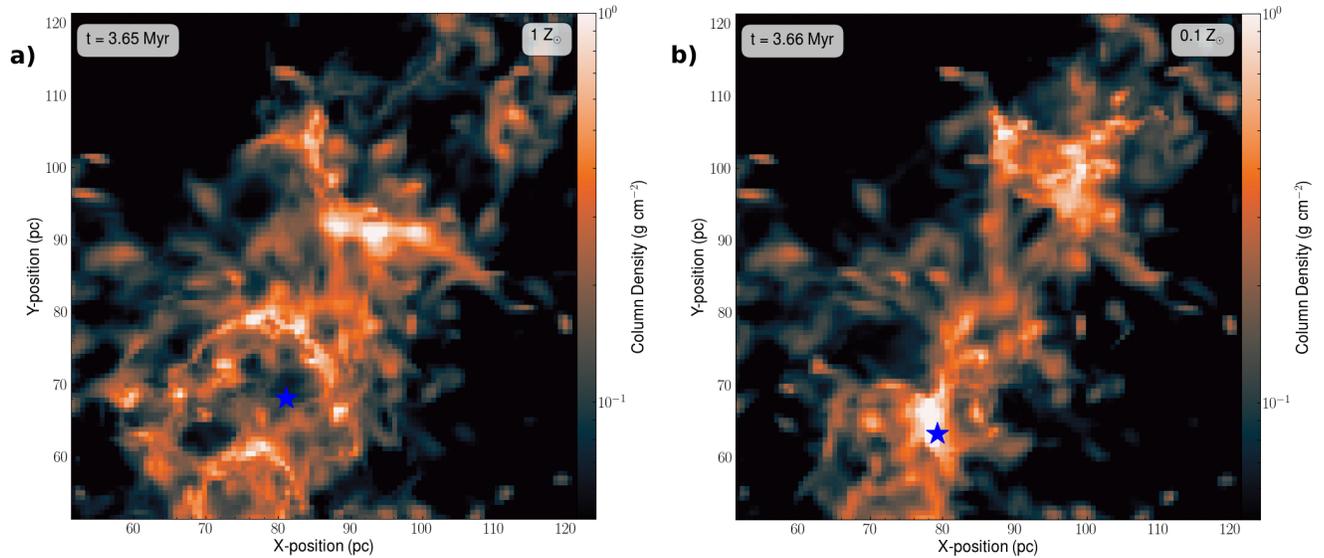

Extended Data Figure 4: **The varying effects of radiative feedback on gas density surrounding YMCs.** (a) The column density along the z-axis near the YMC for the $Z_\odot$ metallicity RHD simulation. The location of the YMC is marked by the blue star. The image is shown just after the YMC begins to clear its surroundings via radiative feedback, resulting in an evacuated bubble at the cluster's location. (b) A plot of the column density in the 0.1 $Z_\odot$ metallicity RHD simulation at the same time as panel (a). The YMC is surrounded by dense gas and there is no radiatively driven bubble.

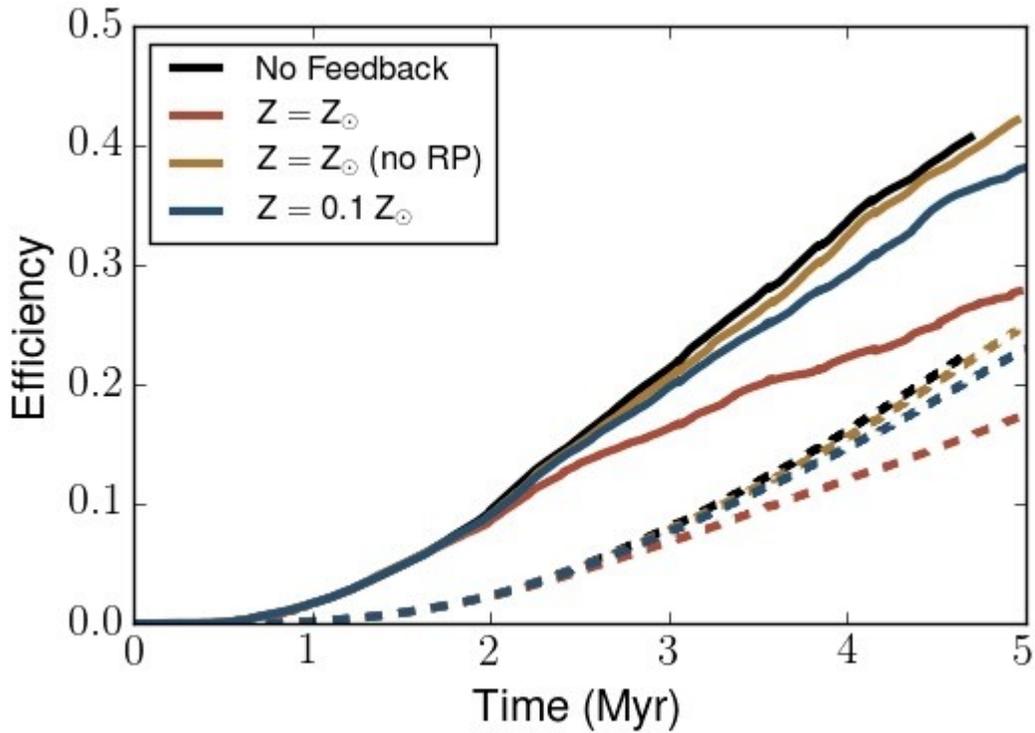

Extended Data Figure 5: **Global cluster particle and star formation efficiency.**
*Solid lines:* The cluster particle formation efficiency (i.e. the total mass in clusters, including stars and gas, divided by the initial GMC mass) plotted versus time.
*Dashed lines:* The star formation efficiency (i.e. the mass of stars only divided by the initial GMC mass). Radiative feedback more strongly affects the $Z_\odot$ simulation. Without radiation pressure (no RP), the efficiencies are nearly indistinguishable from the "no-feedback" case.